\documentclass[12pt]{article}
\usepackage{epsf}
\usepackage{amsmath}
\usepackage{graphicx}
\usepackage{color}
\usepackage{cite}
\usepackage{here}
\usepackage[colorlinks,citecolor=blue]{hyperref}
\renewcommand{\thefootnote}{\#\arabic{footnote}}

\renewcommand{\thefootnote}{\fnsymbol{footnote}}
\def\thefootnote{\fnsymbol{footnote}}

\makeatletter

\@addtoreset{equation}{section}
\makeatother

\begin{document}

\begin{titlepage}

\begin{center}

\vskip .75in

{\Large \bf Variation of the fine structure constant  \vspace{2mm} \\ 
in the light of recent helium abundance measurement} 

\bigskip

\vskip .75in

{\large
Osamu~Seto$\,^1$, Tomo~Takahashi$\,^2$ and Yo~Toda$\,^1$ 
}

\vskip 0.25in

{\em
$^{1}$Department of Physics, Hokkaido University, Sapporo 060-0810, Japan \vspace{2mm} \\
$^{2}$Department of Physics, Saga University, Saga 840-8502, Japan  
}

\end{center}
\vskip .5in

\begin{abstract}

We point out that the recent result of primordial helium-4 ($^4$He) abundance measurement by EMPRESS, which has reported a smaller $^4$He abundance than other measurements, can be well fitted by assuming a time-variation of the fine structure constant~$\alpha$ which is slightly smaller than the present value during big bang nucleosynthesis (BBN). We find that  the EMPRESS result in combination with deuterium abundance measurement indicates $-2.6\% <\Delta\alpha/\alpha <-1.4 \%$ (68\% C.L.) where $\Delta \alpha$ is the difference between the values of $\alpha$ at the BBN and present epochs,  while $-1.2\% <\Delta\alpha/\alpha <0.4 \%$ (68\% C.L.) is obtained from other previous $^4$He abundance data. We also investigate its effects in the framework where the effective number of neutrino species and the lepton asymmetry, which are other typical interpretations of the EMPRESS result, are allowed to vary.  Once a smaller $\alpha$ is adopted, the EMPRESS result can be explained without assuming any non-standard values for the effective number of neutrino species and lepton asymmetry.

\end{abstract}

EPHOU-23-010

\end{titlepage}

\renewcommand{\thepage}{\arabic{page}}
\setcounter{page}{1}
\renewcommand{\thefootnote}{\#\arabic{footnote}}
\setcounter{footnote}{0}

\section{Introduction \label{sec:intro}}

Time variation of the fundamental constants such as the fine structure constant~$\alpha$, the electron mass, the gravitational constant can arise in various theories such as superstring theories, scalar-tensor theories, models with extra dimension (for reviews, see, e.g., \cite{Uzan:2002vq,Uzan:2010pm,Martins:2017yxk}). The variation of such fundamental constants at various cosmic epochs have been probed with a wide range of observations such as quasar absorption spectra \cite{1956Natur.178..688S,Webb:1998cq,Webb:2000mn,Srianand:2004mq}, cosmic microwave background (CMB) \cite{Kaplinghat:1998ry,Avelino:2001nr,Landau:2001st}, big bang nucleosynethesis (BBN) \cite{Bergstrom:1999wm,Ichikawa:2002bt,Muller:2004gu,Coc:2006sx,Nollett:2002da,Ichikawa:2004ju}, the dating of meteorites \cite{1958PMag....3..582W,Olive:2002tz,Olive:2003sq}, most of which put severe constraints on them~\cite{Uzan:2002vq,Uzan:2010pm,Martins:2017yxk}. Some studies have obtained non-null results for the variation of some constants in the past from the natural reactor such as at Oklo \cite{Fujii:2002hc} and quasar absorption spectra~\cite{Webb:1998cq,Murphy:2003hw}, those have been debated for example in Refs.~\cite{Gould:2006qxs,Bonifacio:2013vfa} nevertheless. Interestingly, the time variation of some fundamental constants has recently been suggested as a possible solution for the so-called Hubble tension ($H_0$ tension), which is the issue of inconsistency (at almost $5 \sigma$ level) between the values of the Hubble constant $H_0$ obtained from local direct measurements such as the Cepheid calibrated supernova distance ladder \cite{Riess:2021jrx} and from indirect measurements such as CMB from Planck satellite \cite{Planck:2018vyg}  (see reviews~\cite{DiValentino:2021izs,Perivolaropoulos:2021jda} for the current status of the tension). For instance, a model with time-varying electron mass has been proposed to reduce the tension significantly \cite{Sekiguchi:2020teg} (see also Refs.~\cite{Planck:2014ylh,Hart:2019dxi,Fung:2021wbz,Hoshiya:2022ady,Seto:2022xgx})\footnote{
It may be worth noting that other cosmological aspects can also be affected such as cosmological constraints on neutrino masses \cite{Sekiguchi:2020igz} in time-varying electron mass model.
} and a rapid transition of the gravitational constant also has been suggested to resolve the $H_0$ tension \cite{Marra:2021fvf}.  

Recently, another cosmological tension seems to appear in the primordial helium-4 abundance $Y_p$, in which the value of $Y_p$ reported by EMPRESS, $Y_p = 0.2370^{+0.0033}_{-0.0034}$~\cite{Matsumoto:2022tlr},  is somewhat smaller than that obtained in other recent observations \cite{Kurichin:2021ppm,Hsyu:2020uqb,Aver:2015iza}. Intriguingly, the combined analysis of the value of $Y_p$ by EMPRESS and the deuterium (D) abundance $D_p$ \cite{Cooke:2017cwo} indicates the effective number of neutrino species $N_{\rm eff}$ deviated from the standard value of $N_{\rm eff} = 3.046$\footnote{
For recent precise calculations of $N_{\rm eff}$ for the standard case, see \cite{deSalas:2016ztq,EscuderoAbenza:2020cmq,Akita:2020szl,Froustey:2020mcq,Bennett:2020zkv}.
} and also a non-zero lepton asymmetry, characterized by the electron neutrino chemical potential or the degeneracy parameter $\xi_e$~\cite{Matsumoto:2022tlr}. The EMPRESS result has stimulated several works for its interpretations and implications of lepton asymmetry~\cite{Kawasaki:2022hvx,Burns:2022hkq,Borah:2022uos,Escudero:2022okz}, modified gravity~\cite{Kohri:2022vst}, early dark energy~\cite{Takahashi:2022cpn}, and so on.

In this paper, we argue that the time variation of $\alpha$ can explain the EMPRESS $Y_p$ result without resorting to non-standard values of $N_{\rm eff}$ and $\xi_e$. To this end, we investigate the light element abundances such as ${}^4$He and D and perform the statistical analysis by using the EMPRESS $Y_p$ result along with the data of D abundance in Ref.~\cite{Cooke:2017cwo}\footnote{
The compatibility of the time-varying $\alpha$ model with other recent BBN observations are discussed~\cite{Deal:2021kjs,Martins:2020syb}.
}.  Indeed, as we will show, $\alpha$ a few percent smaller than the present value is preferred by the EMPRESS result and in such a case, the standard values of $N_{\rm eff} = 3.046$ and $\xi_e=0$ are well within the allowed region.  

The structure of this paper is as follows. In the next section, we summarize how the time variation of $\alpha$ is implemented in the BBN calculation to predict the light abundances and the method of our statistical analysis. In the Section~\ref{sec:results}, we present our results: the time variation of $\alpha$ is indeed preferred by the recent EMPRESS $Y_p$ result and also we do not need to assume any other non-standard cosmological scenario. We conclude this paper in the final section.


\section{Setup and analysis method \label{sec:setup}}


In this section, we summarize the methodology of our analysis. First, we describe how we implement the time-varying fine structure constant to calculate the abundances of light elements. The description of our statistical analysis methods follows.


\subsection{Effects of varying fine structure constants}


We apply the following modifications to implement the effects of varying fine structure constant. For the details of what aspects of BBN calculation are modified due to the change of $\alpha$, see e.g., Refs.~\cite{Bergstrom:1999wm,Nollett:2002da,Ichikawa:2002bt,Ichikawa:2004ju}.

\subsubsection*{Weak reaction rate correction due to the mass difference variation }

Increasing $\alpha$ decreases the mass
difference between neutrons and protons ($Q\equiv m_{n}-m_{p}$).  The response of the variation of $\alpha$  on $Q$ is given by \cite{Gasser:1982ap} 
\begin{equation}
Q=\left(-0.76\frac{\alpha}{\alpha_{0}}+2.05 \right)\,\mathrm{MeV} ,
\end{equation}
with the present value of the fine structure constant $\alpha_0 \simeq 1/137.036$.

The change of $Q$ is implemented in the weak reaction rate in the
BBN era as 
\begin{equation}
\Gamma(n\rightarrow p^{+}e^{-}\nu_{e})_{\mathrm{BBN}}=\Gamma_{n0}\frac{\lambda_{n\rightarrow p^{+}}(Q)}{\lambda_{n\rightarrow p^{+}}(Q_{0})},
\end{equation}
where $\Gamma_{n0} = 1/ \tau_{n0} \simeq 1/(879.4~{\rm s})$  is the present values of the
neutron decay rate, $Q_{0}\simeq 1.29~\mathrm{MeV}$ is the present value of
the mass difference and $\lambda$ is the phase space integral~\cite{Weinberg:1972kfs}
\begin{equation}
\lambda_{n\rightarrow p^{+}}(Q)\propto\left(\int_{-\infty}^{-m_{e}-Q}\frac{x^{2}(Q+x)^{2}\sqrt{1-\frac{m_{e}^{2}}{(Q+x)^{2}}}}{\left(e^{\frac{x}{T_{\nu}}}+1\right)\left(e^{-\frac{Q+x}{T}}+1\right)}dx+\int_{m_{e}-Q}^{\infty}\frac{x^{2}(Q+x)^{2}\sqrt{1-\frac{m_{e}^{2}}{(Q+x)^{2}}}}{\left(e^{\frac{x}{T_{\nu}}}+1\right)\left(e^{-\frac{Q+x}{T}}+1\right)}dx\right),\label{Eq:phaseint}
\end{equation}
with $T_{\nu}$ being the temperature of neutrinos, and a similar expression holds for the $p^{+}\rightarrow n$ weak interaction rate.

Although we assume that $\alpha$ does not change during
the BBN epoch, it should eventually evolve to the present value at some
point in the course of the evolution of the Universe. We have implemented
the above change in \texttt{PArthENoPE} \cite{Pisanti:2007hk,Consiglio:2017pot,Gariazzo:2021iiu},
which we use to calculate the abundances of light elements such as
D and ${}^4$He. 

\subsubsection*{Reaction rates corrections}

Effects of the time-variation of $\alpha$ also appears in the Coulomb barrier penetrability. For charged-particle
reactions, the expression of the cross section $\sigma(E)$ is given
by

\begin{equation}
\sigma(E)=\frac{S(E)}{E}\exp(-2\pi\alpha Z_{i}Z_{j}\sqrt{\mu/2E}),
\end{equation}
where $S(E)$ is the astrophysical $S$-factor, $\mu$ is the reduced
mass, and $Z_{i,j}$ are the atomic number of the nuclei. Since a smaller
$\alpha$ increases the cross section, the nuclear reactions proceed
more quickly.  Following Ref.~\cite{Bergstrom:1999wm}, the $\alpha$ dependence of the reaction rates are included by modifying \texttt{PArthENoPE3.0}.

\subsubsection*{Other corrections}

We have also taken account of other minor effects~\cite{Nollett:2002da}: 
final-state Coulomb interaction corrections, radiative capture corrections, mass corrections due to the electromagnetic contributions $\Delta M_{EM}$, and radiative capture corrections.

\subsubsection*{Results}
\begin{figure}[ht]
\begin{centering}
\includegraphics[width=8cm]{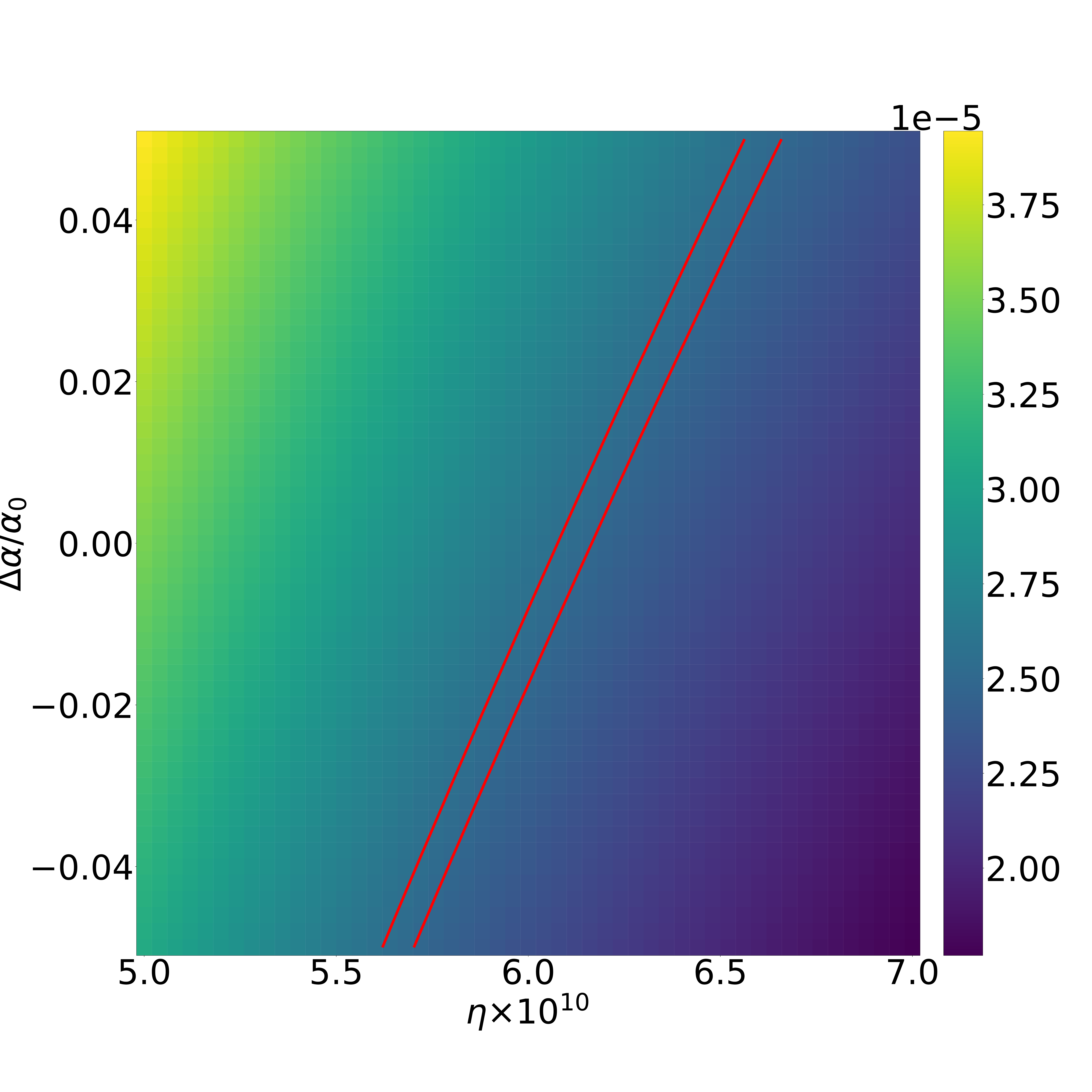}\hspace{2mm}\includegraphics[width=8cm]{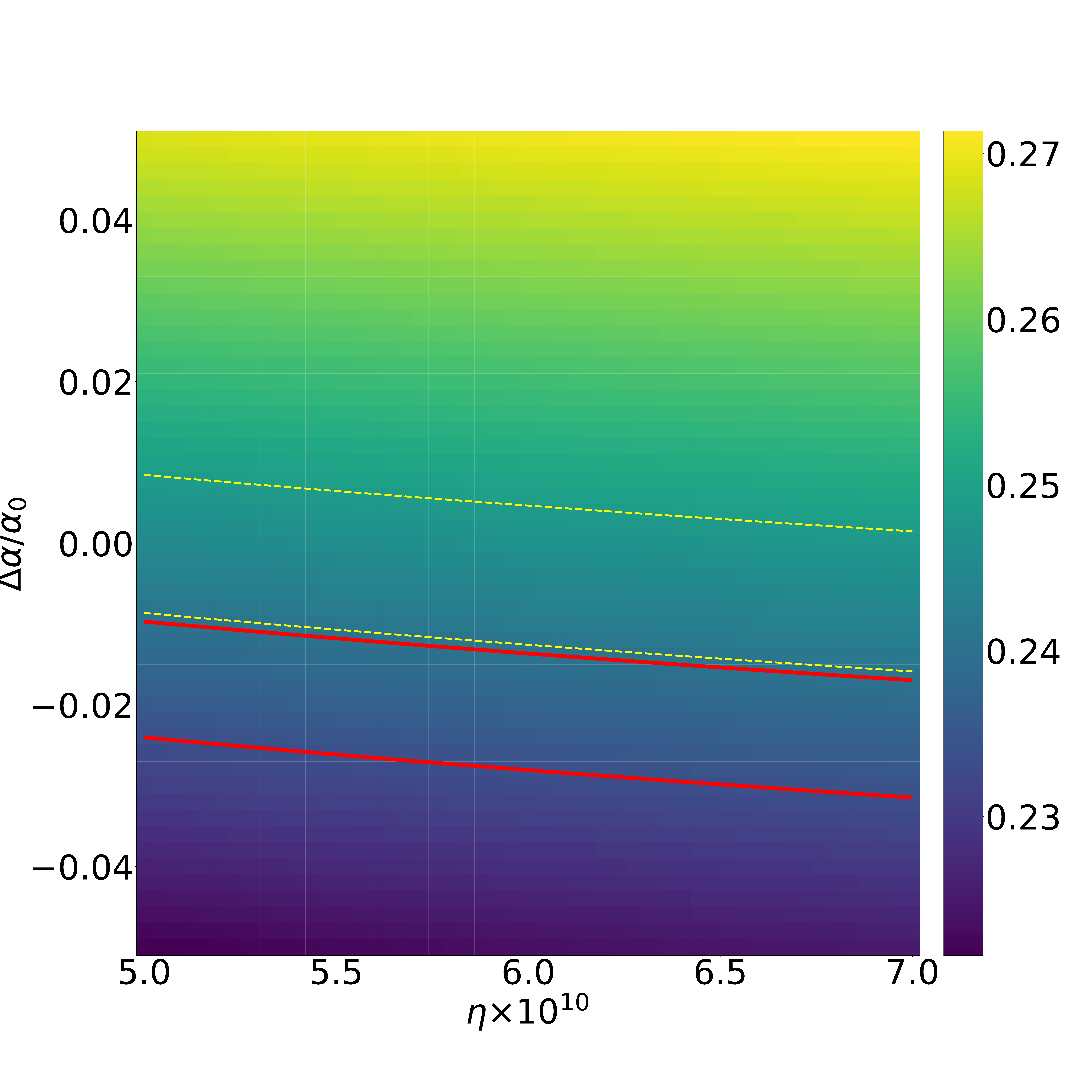}\\
 
\par\end{centering}
\caption{\label{fig:D_He} Density plots of $D_{p}$ (left) and $Y_{p}$ (right)
in the $\eta$ -- $\Delta \alpha/\alpha_{0}$ plane. In this figure, other
parameters are fixed as $\Delta N_{{\rm eff}}=0$ (the deviation from the standard value of $N_{\rm eff}$) and $\xi_{e}=0$.
Regions between red lines in the left and right panels correspond
to 1$\sigma$ allowed region from measurements of $D_{p}$ \cite{Cooke:2017cwo}
and the EMPRESS $Y_{p}$ \cite{Matsumoto:2022tlr}, respectively. 
Regions between green dashed lines in the right panel correspond
to 1$\sigma$ allowed region from  another $Y_p$ abundance measurement by Aver et al.~\cite{Aver:2015iza}. 
}
\end{figure}

In Fig.~\ref{fig:D_He}, we show density plots of $D_{p}$ (left)
and $Y_{p}$ (right) in the $\eta$ -- $\Delta \alpha/\alpha_{0}$ plane
where $\eta$ is the baryon-to-photon ratio and $\Delta \alpha = \alpha - \alpha_0$. Regions between red lines
in the left and right panels correspond to 1$\sigma$ allowed region
from measurements of $D_{p}$ \cite{Cooke:2017cwo} and the EMPRESS
$Y_{p}$ \cite{Matsumoto:2022tlr}, respectively. 
For comparison, in the right panel, another recent helium abundance measurement from Aver et al.~\cite{Aver:2015iza} is also shown as region between green dashed lines. 
As is shown in the figure, as the structure constant $\alpha$ decreases, the abundance of ${}^4$He become smaller. Therefore the value of $Y_{p}$ reported by EMPRESS can be explained by a smaller structure constant. The abundance of D is also affected by the change of $\alpha$, however, its effect is canceled by the change of $\eta$ as we explain below.

The response of $Y_{p}$ against the change of $\alpha$ can be understood
by noticing that the mass difference $Q$ decreases by increasing $\alpha$ and
 the neutron decay rate becomes larger (the neutron lifetime becomes shorter), which makes 
the abundances of ${}^4$He larger. These behaviors in response to the variation of the parameters, denoted as $\Delta \alpha$ and $\Delta\eta$, can be understood by the following scaling relation \cite{Dent:2007zu}: 
\begin{eqnarray}
\frac{\Delta D_p}{D_p} &=&3.6 \frac{\Delta \alpha}{\alpha}-1.6\frac{\Delta \eta}{\eta} , \\ [8pt]
\frac{\Delta Y_p}{Y_p} & =&1.9 \frac{\Delta \alpha}{\alpha}+0.04\frac{\Delta \eta}{\eta} .
\end{eqnarray}
The $\alpha$ dependence of $D_p$ can be somewhat canceled by the change of $\eta$. This behavior indeed can be seen in the left panel of Fig.~\ref{fig:D_He}. On the other hand, the ${}^4$He abundance is not so sensitive with respect to the variation of $\eta$, and hence $\alpha$ almost controls the value of $Y_p$. This also can be observed in the right panel of Fig.~\ref{fig:D_He}.

\subsection{Analysis method}

In the following, we investigate the impact of the EMPRESS result
of $Y_{p}$ on models beyond the standard assumption, particularly
focusing on the time-varying $\alpha$. Since it has been discussed
that the EMPRESS result may suggest non-standard values of the effective
number of neutrino $N_{{\rm eff}}$ and non-zero lepton asymmetry,
characterized by the chemical potential (the degeneracy parameter)
of the electron neutrino $\xi_{e}$ \cite{Matsumoto:2022tlr}, we
also vary $N_{{\rm eff}}$ and $\xi_{e}$ in models with varying $\alpha$
in some cases.

To study the preferred ranges for $\alpha,N_{{\rm eff}}$ and $\xi_{e}$,
we calculate the value of $\chi^{2}$ fitted to the observed values
of $Y_{p}$ by EMPRESS \cite{Matsumoto:2022tlr} and $D_{p}$~\cite{Cooke:2017cwo}. 
We also adopt the prior for the baryon density
derived from the CMB measurement of Planck \cite{Planck:2018vyg}.
The total value of $\chi^{2}$ is 
evaluated as
\begin{equation}
\chi^{2}(\alpha,\eta,\xi_{e},N_{{\rm eff}})=\chi_{Y_{p}}^{2}(\alpha,\eta,\xi_{e},N_{{\rm eff}})+\chi_{D_{p}}^{2}(\alpha,\eta,\xi_{e},N_{{\rm eff}})+\chi_{\eta}^{2}(\eta) \,,
\label{eq:chi2} 
\end{equation}
where $\chi_{Y_{p}}^{2}$ and $\chi_{D_{p}}^{2}$ are 
the one calculated from the fit to the observed $Y_{p}$ by EMPRESS \cite{Matsumoto:2022tlr}
and the measurement of $D_{p}$~\cite{Cooke:2017cwo}, which are respectively given as 
\begin{eqnarray}
 &  & \chi_{Y_{p}}^{2}(\alpha,\eta,\xi_{e},N_{{\rm eff}})=\frac{(Y_{p,{\rm obs}}-Y_{p}^{{\rm (th)}}(\alpha,\eta,\xi_{e},N_{{\rm eff}}))^{2}}{\sigma_{Y_{p},{\rm obs}}^{2}+\sigma_{Y_{p},{\rm th}}^{2}}\,,\label{eq:chi2_Yp}\\
 &  & \chi_{D_{p}}^{2}(\alpha,\eta,\xi_{e},N_{{\rm eff}})=\frac{(D_{p,{\rm obs}}-D_{p}^{{\rm (th)}}(\alpha,\eta,\xi_{e},N_{{\rm eff}}))^{2}}{\sigma_{D_{p},{\rm obs}}^{2}+\sigma_{D_{p},{\rm th}}^{2}}\,,\label{eq:chi2_Dp}
\end{eqnarray}
where $Y_{p,{\rm obs}}=0.2370$ and $\sigma_{Y_{P},{\rm obs}}=0.00335$
are the mean value and 1$\sigma$ error from the observation of $Y_{p}$
\cite{Matsumoto:2022tlr}. For comparison, we also make an analysis using the measurement of $Y_p$ from Aver et al.~\cite{Aver:2015iza} $Y_p = 0.2449 \pm 0.040 $ in some cases. For deuterium abundance, we take $D_{p,{\rm obs}}=2.527\times10^{-5}$ and $\sigma_{D_{p,{\rm obs}}}=0.030\times10^{-5}$ which are from  Cooke et al.~\cite{Cooke:2017cwo}. 
$\sigma_{Y_{p},{\rm th}}=\sqrt{0.00003^{2}+0.00012^{2}}$ is the theoretical
uncertainty originating from the nuclear rates  and the neutron lifetime, respectively corresponding to the first and second terms in the square root, 
and $\sigma_{D_{p},{\rm th}}=\sqrt{0.06^{2} \times 10^{-10}}$ is the one for $D_p$ from the nuclear reaction rates~\cite{Gariazzo:2021iiu}.

$Y_{p}^{{\rm (th)}}(\alpha,\eta,\xi_{e},N_{{\rm eff}})$ and $D_{p}^{{\rm (th)}}(\alpha,\eta,\xi_{e},N_{{\rm eff}})$
are theoretically predicted values of $Y_{p}$ and $D_{p}$ calculated
by using \texttt{PArthENoPE} \cite{Pisanti:2007hk,Consiglio:2017pot,Gariazzo:2021iiu} which we have modified to implement the variation of $\alpha$. 
$\chi_{\eta}^{2}$ is to take account of the prior on the baryon density
from the CMB measurement of Planck, which is written as 
\begin{equation}
\chi_{\eta}^{2}(\eta)=\frac{(\eta_{{\rm Planck}}-\eta)^{2}}{\sigma_{\eta_{{\rm Planck}}}^{2}} , 
\label{eq:chi2_eta}
\end{equation}
where $\eta_{{\rm Planck}}=6.082\times10^{-10}$ and $\sigma_{\eta_{\rm Planck}}=0.060\times10^{-10}$, 
{which is taken from the analysis for the case of $\Lambda$CDM$+N_{\rm eff}+Y_p$ model \cite{Planck:2018vyg}. 
In the following analysis, we also vary the value of $N_{\rm eff}$ and $\xi_e$, and hence we take the value of baryon density from such an analysis\footnote{
Since the analysis varying $\xi_e$ is not available in Ref.~\cite{Planck:2018vyg}, we instead adopt the constraint on the baryon density in the case of $\Lambda$CDM$+N_{\rm eff}+Y_p$ model, as has also been done in Ref.~\cite{Matsumoto:2022tlr}.
}.  Since $N_{\rm eff}$ and $Y_p$ are correlated with the baryon density in CMB (see, e.g., \cite{Planck:2018vyg,Ichikawa:2006dt,Ichikawa:2007js,Ichikawa:2008pz}), the uncertainty of $\eta$ is somewhat larger than that for the $\Lambda$CDM case.

We calculate $\chi^{2}$ given in Eq.~\eqref{eq:chi2} to obtain
constraints on $\alpha,N_{{\rm eff}},\xi_{e}$ and $\eta$. We mainly
show our results in a two dimensional plane of two of these parameters.
We present constraints on a two dimensional plane by marginalizing over
the other parameter(s), which in practice is done by determining (fixing)
the redundant parameter to the one minimizing the value of $\chi^{2}$.
In some analyses, we use $\eta$ as one of variables to show a two dimensional plane and  we do not include $\chi_{\eta}^{2}$ in the total $\chi^{2}$.

\section{Results \label{sec:results}}

\begin{figure}[t]
\begin{centering}
\includegraphics[width=10cm]{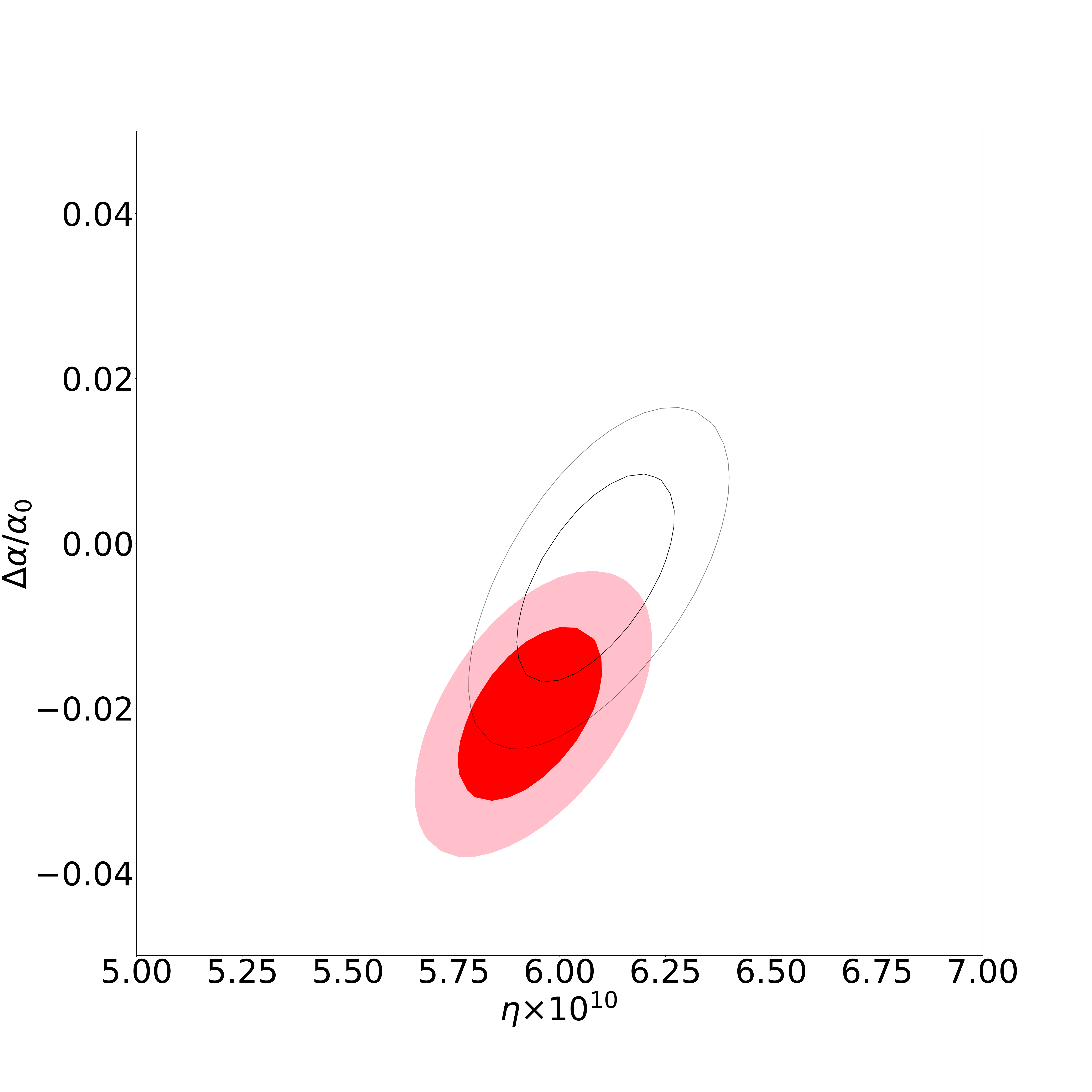} 
\par\end{centering}
\caption{\label{fig:alpha_chi2} 1$\sigma$ and 2$\sigma$ constraints in the $\eta$--$\Delta \alpha/\alpha_{0}$
plane from the EMPRESS measurements of $Y_{p}$ and $D_{p}$ (red shaded). 
We also show the case where we adopt the $Y_{p}$ measurement by Aver et al. and $D_{p}$ (unshaded black contours). 
In this figure, other parameters are fixed as $\Delta N_{{\rm eff}}=0$ and $\xi_{e}=0$.  In this figure, $\chi_\eta^2$ is not included in the total $\chi^2$.}
\end{figure}

\begin{figure}[h]
\begin{centering}
\includegraphics[width=10cm]{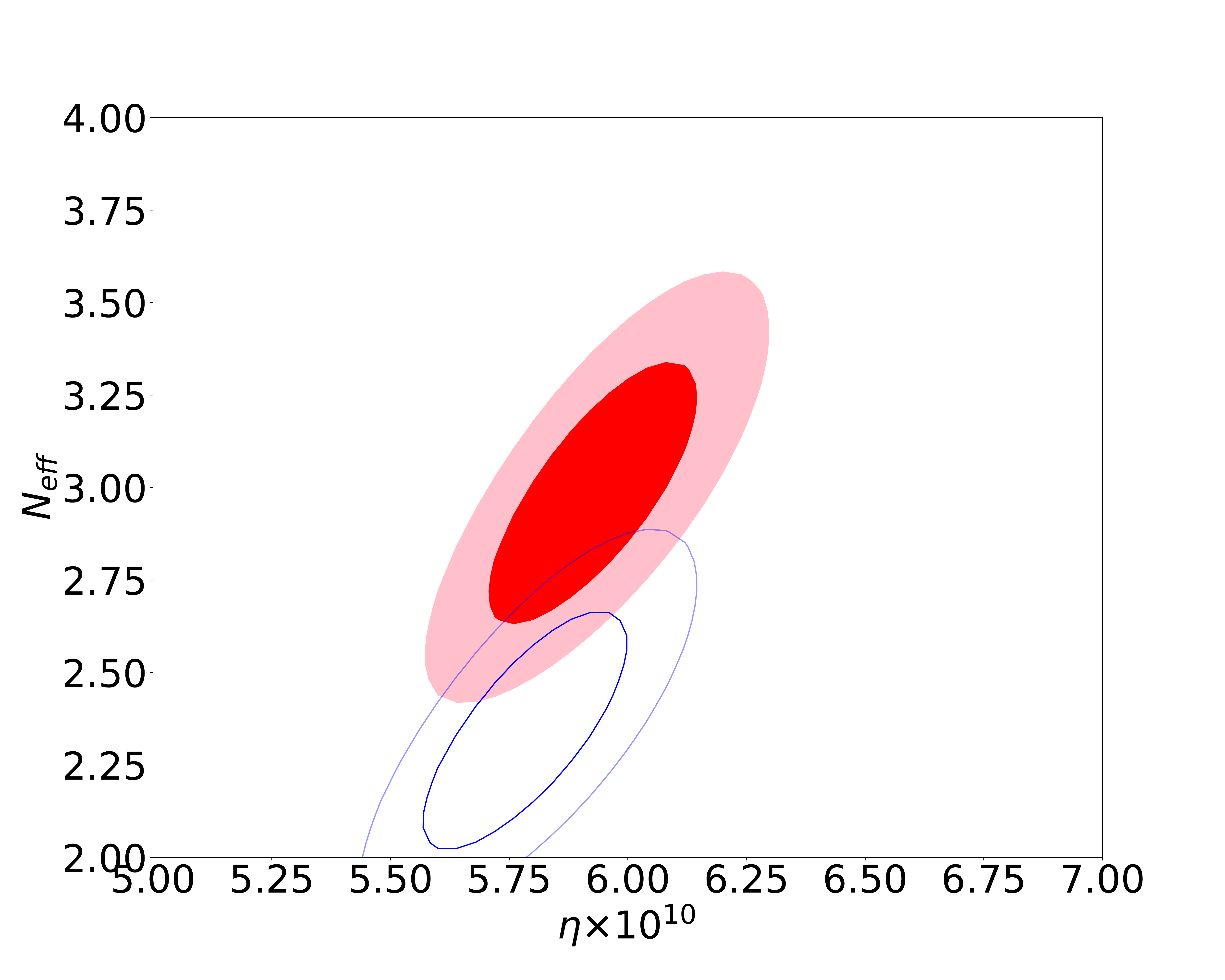} 
\par\end{centering}
\caption{\label{fig:eta_Neff}  1$\sigma$ and 2$\sigma$ constraints in the $\eta$--$N_{{\rm eff}}$
plane from the measurements of $Y_{p}$ and $D_{p}$ with the electron
mass being fixed as $\alpha/\alpha_{0}=0.98$ (red shaded). The neutrino
chemical potential is set to $\xi_{e}=0$. For reference, the constraint
for the standard case (i.e., $\alpha/\alpha_{0}=1$) is shown with
blue lines. In this figure, $\chi_\eta^2$ is not included in the total $\chi^2$.}
\end{figure}

Now in this section, we present our results by showing constraints
on the parameters obtained by fitting to the data of $Y_{p}$  from EMPRESS~\cite{Matsumoto:2022tlr} and $D_{p}$ from Cooke et al.~\cite{Cooke:2017cwo} using $\chi^{2}$ explained in the previous section. First, we
show the constraint shaded with red in the $\eta$--$\Delta \alpha/\alpha_{0}$ plane in
Fig.~\ref{fig:alpha_chi2}, in which we fix $\Delta N_{{\rm eff}}$
and $\xi_{e}$ to be the standard values, i.e., $\Delta N_{{\rm eff}}=0$
and $\xi_{e}=0$. 
For comparison, we also show the constraints for the case of the $Y_{p}$ measurement by Aver et al.~\cite{Aver:2015iza} with the same $D_p$ data with unshaded black contours. 
The figure indicates that the best-fit value for the case with EMPRESS+Cooke et al. is $\alpha/\alpha_{0}\simeq0.98$, and $\alpha/\alpha_{0}\simeq 1$
is excluded more than 2$\sigma$, which shows that the fine structure
constant should be decreased by 2 \% during BBN epoch compared to
the present value when the EMPRESS $Y_p$ is adopted, on the other hand, $Y_p$ from Aver et al. shows no such deviation.  In any case,  the varying $\alpha$ scenario seems to
be favored from the recent EMPRESS result of $Y_{p}$. After marginalizing over $\eta$,   the bound for $\alpha$ is obtained as  $-2.6\% <\Delta\alpha/\alpha_0 <-1.4 \% \, (1\sigma)$ (EMPRESS+Cooke et al.) 
and $-1.2\% <\Delta\alpha/\alpha_0 <0.4 \% \, (1\sigma)$ (Aver et al.+Cooke et al.).

Next, we show the constraint in the $\eta$--$N_{{\rm eff}}$ plane
in Fig.~\ref{fig:eta_Neff}. Here we fix $\alpha/\alpha_{0}=0.98$
(red shaded region), which is motivated from the result presented
in Fig.~\ref{fig:alpha_chi2}, and $\alpha/\alpha_{0}=1$ (blue contours),
which is the standard case, for comparison.  Indeed it has already been pointed out that the EMPRESS $Y_p$ result  favors $N_{{\rm eff}}<3.046$  when there is no lepton asymmetry (i.e., $\xi_{e}=0$), and the baryon
density is also deviated from that obtained by Planck~\cite{Matsumoto:2022tlr}, although the discrepancy is not so significant. 
On the other hand, as seen from Fig.~\ref{fig:eta_Neff}, when we assume $\alpha/\alpha_{0}=0.98$, the
standard value of $\Delta N_{{\rm eff}}=0$ is well within 1$\sigma$ allowed region, and moreover, the baryon density can also be consistent
with that obtained from Planck ($\eta_{{\rm Planck}} \simeq 6.1\times10^{-10}$).
Therefore, the time-varying fine structure constant can remove the tensions between the EMPRESS $Y_{p}$ result and the standard cosmological
model.
\begin{figure}[ht]
\includegraphics[width=8.1cm]{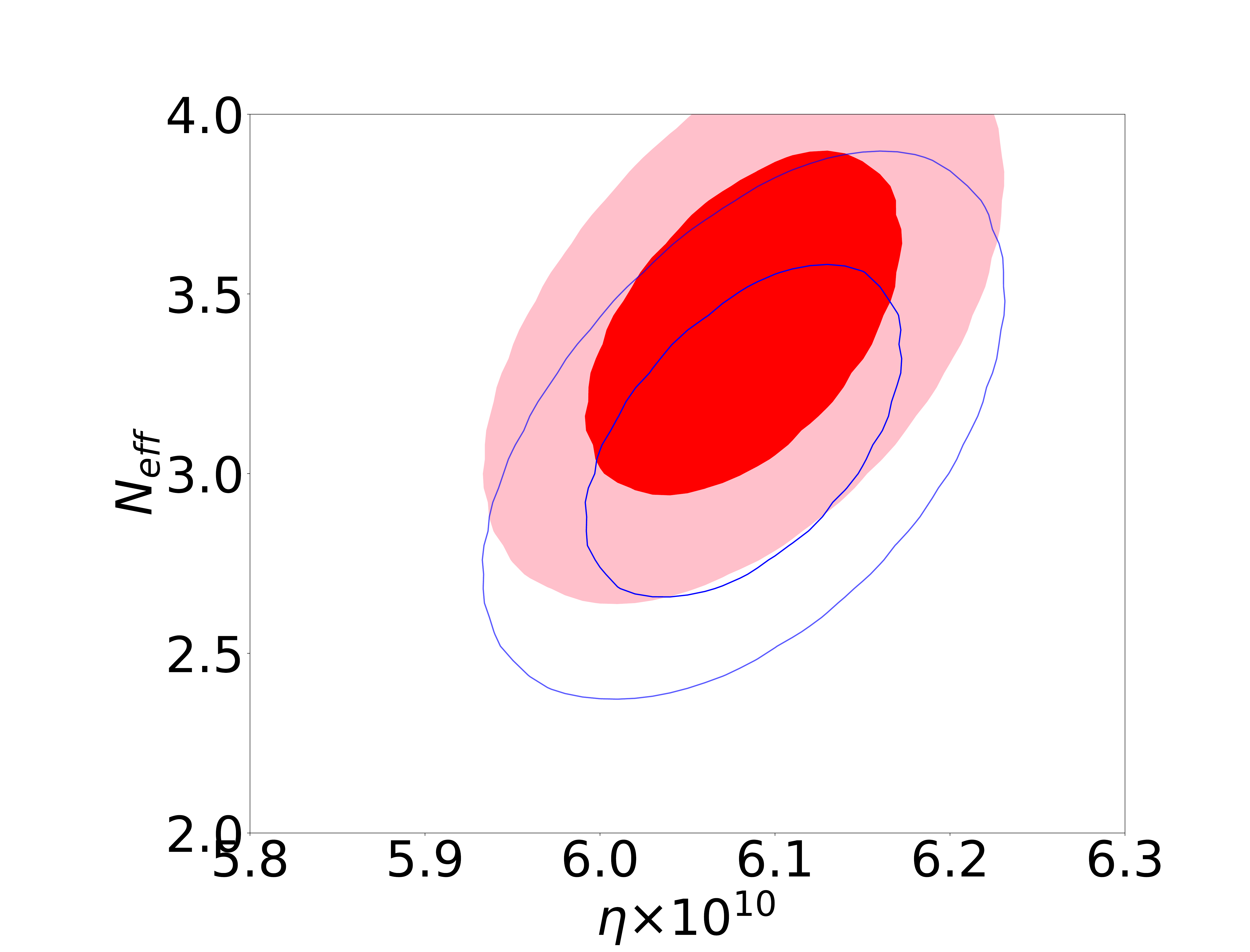}

\includegraphics[width=8.1cm]{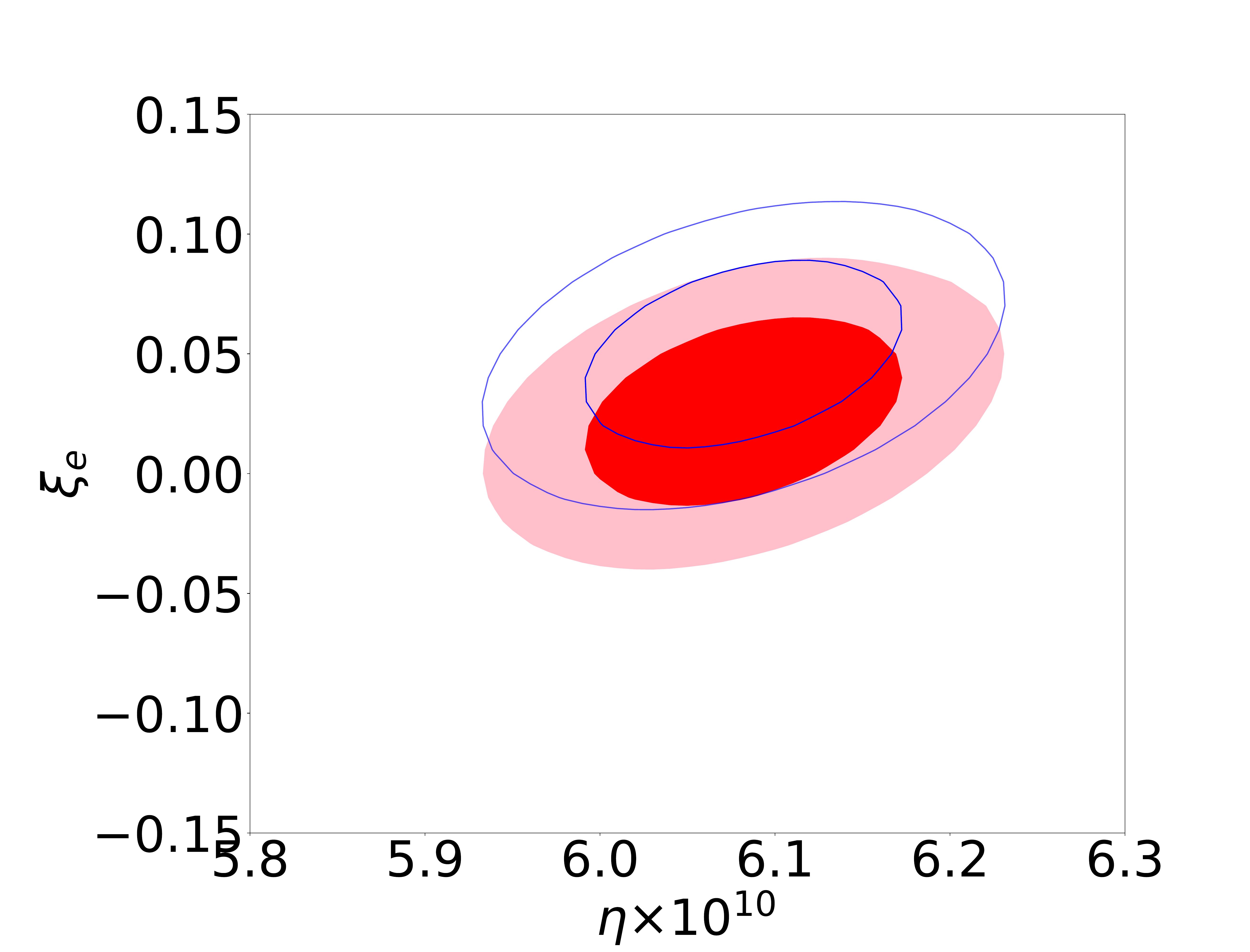}
\includegraphics[width=8.1cm]{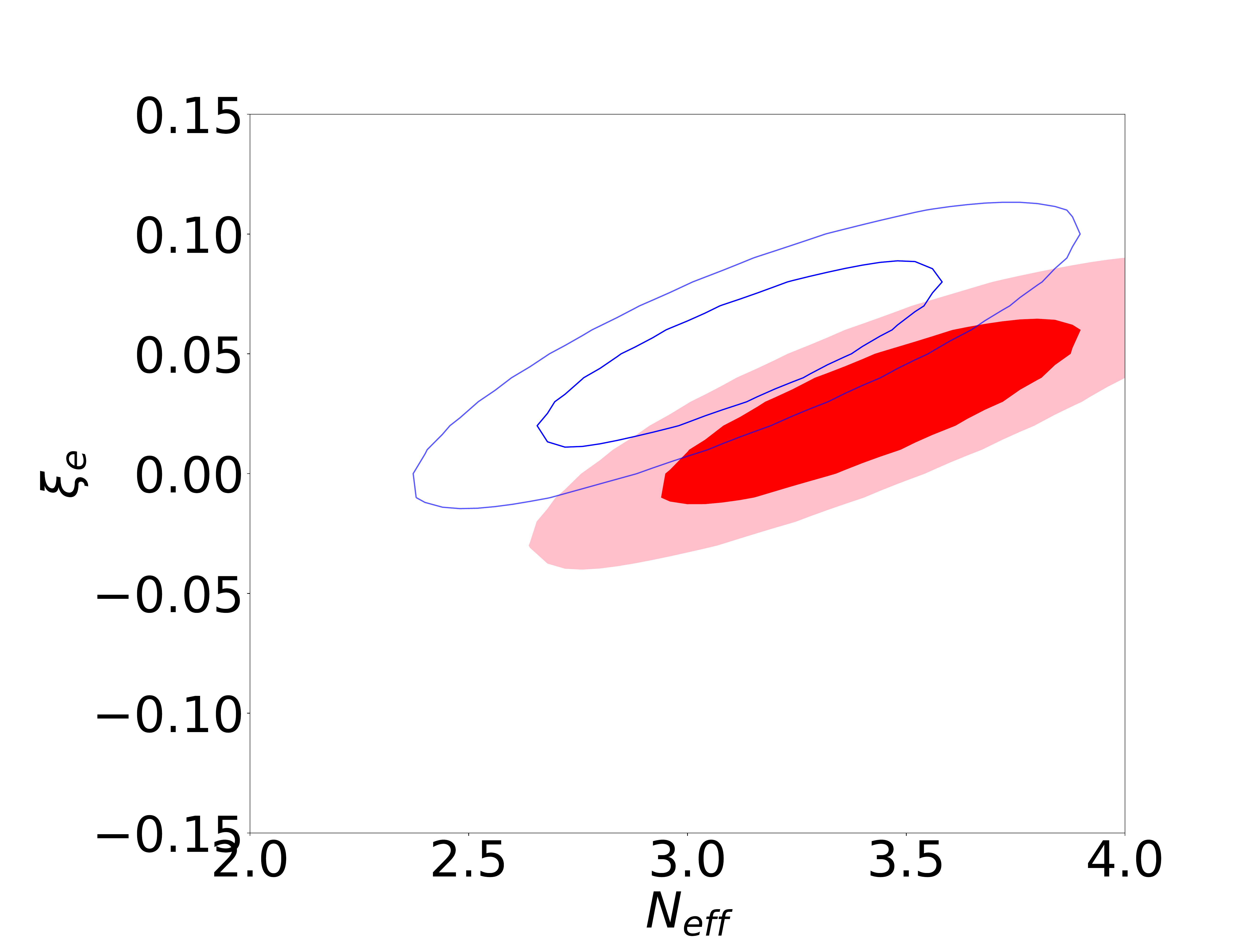} 
\par
\caption{\label{fig:Neff_xie_eta}  1$\sigma$ and 2$\sigma$ constraints on the $\eta$--$N_{{\rm eff}}$
plane (top), the $N_{{\rm eff}}$--$\xi_{e}$ plane (bottom right) and
the $\eta$--$\xi_{e}$ plane (bottom left). Here we fix the structure constant
as $\alpha/\alpha_{0}=0.98$ (red shaded region) and $1$ (blue contours), but vary 3 other parameters: $N_{{\rm eff}},\xi_{e}$
and $\eta$. To project the constraints on 2 dimensional plane, we
have eliminated the redundant parameter which is not shown in the
plane by fixing it so that the value of $\chi^{2}$ is minimized for
the parameter. }              
\end{figure}

\begin{figure}[ht]
\begin{centering}
\includegraphics[width=8cm]{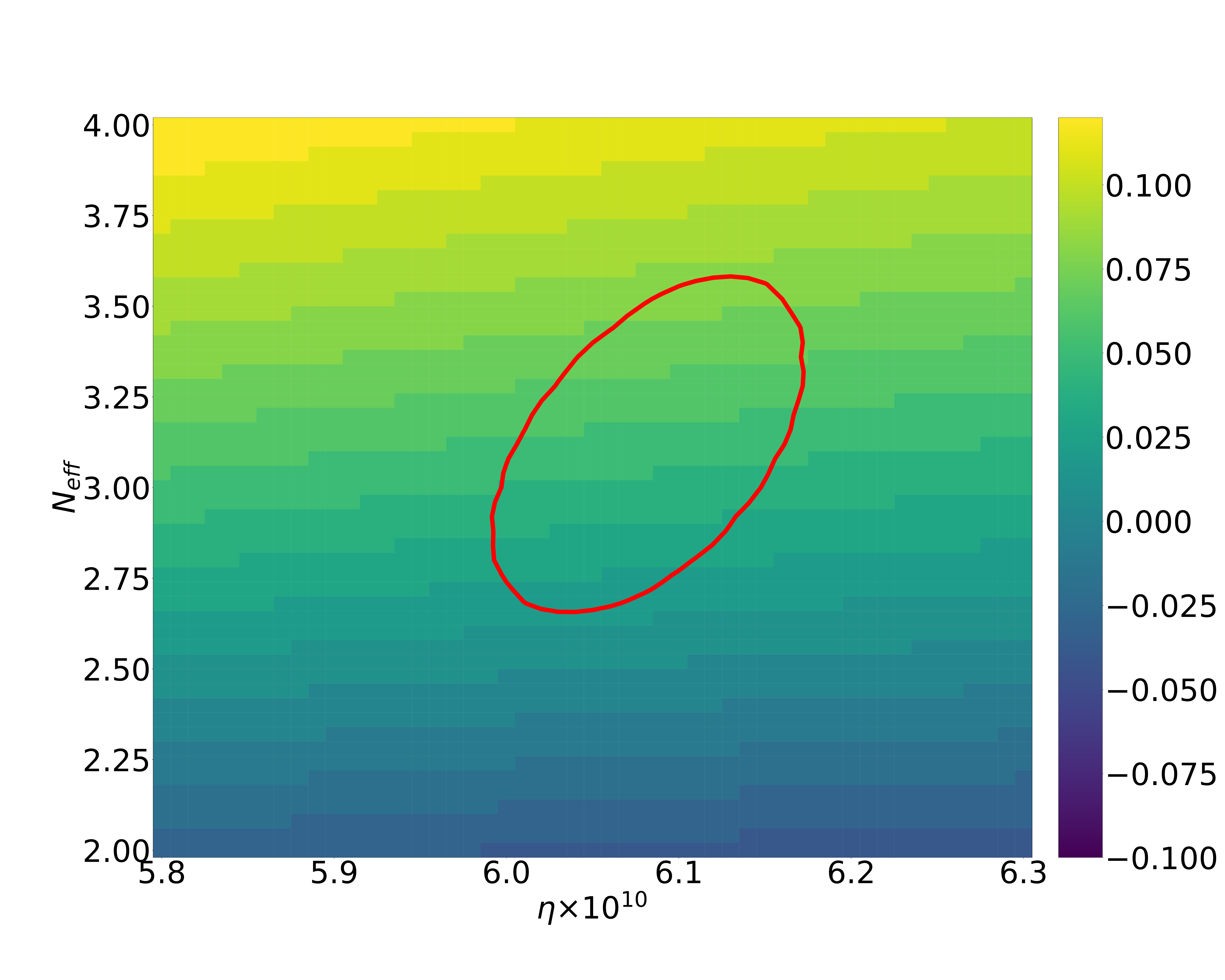}
\includegraphics[width=8cm]{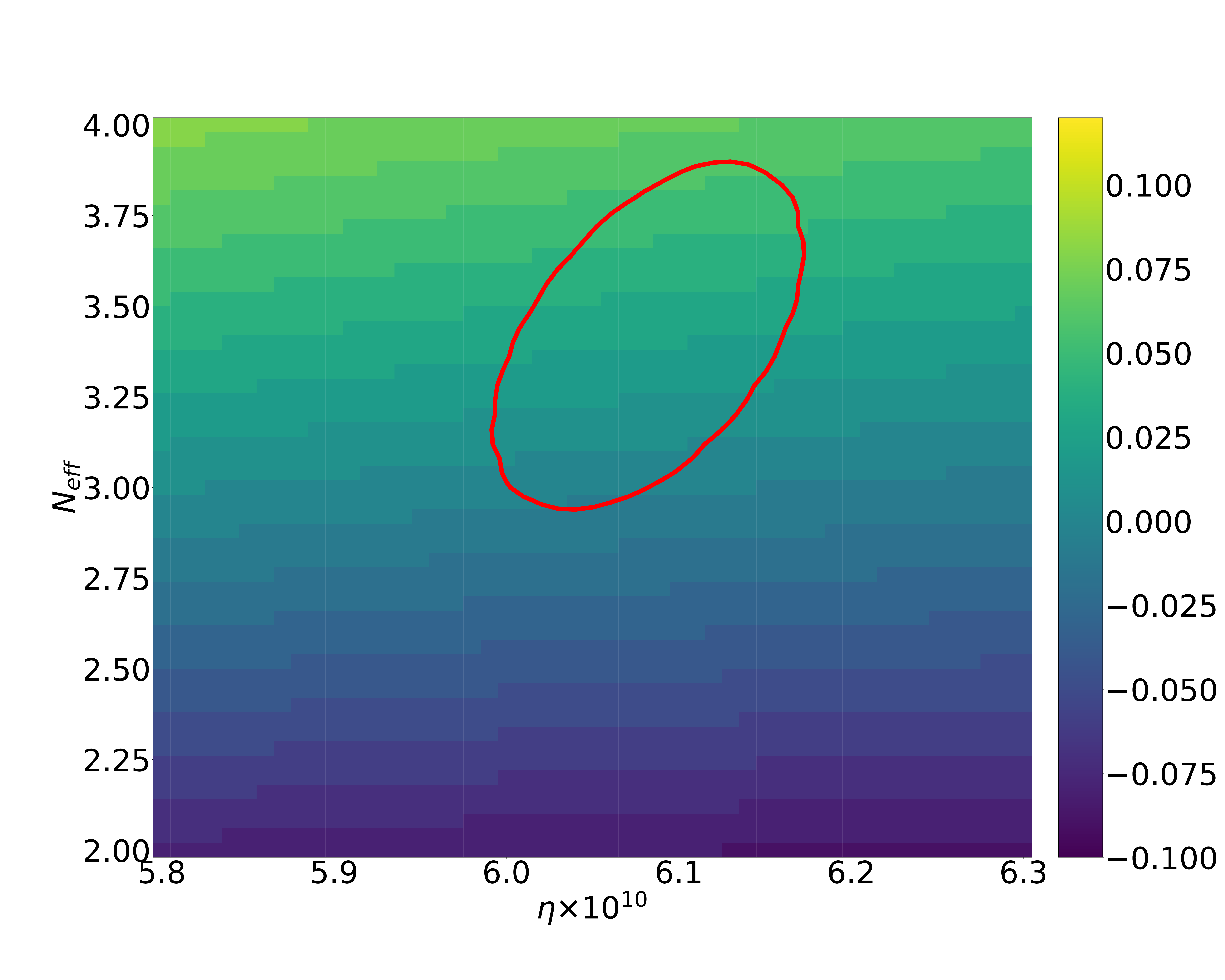}
\par\end{centering}
\caption{\label{fig:Neff_eta_byxie} Density plots of $\xi_e$ which minimizes $\chi^2$ in $\eta$--$N_{{\rm eff}}$ plane for the cases with $\alpha/\alpha_{0}=1$ (left panel) and $\alpha/\alpha_{0}=0.98$ (right panel).
Ellipse surrounded by red lines correspond to 1$\sigma$ allowed region in Fig.~\ref{fig:Neff_xie_eta}.
}
\end{figure}

To see how the preference for small $\alpha$ by the EMPRESS value
of $Y_{p}$ is affected by the existence of the lepton asymmetry characterized
by $\xi_{e}$, we also analyze the cases where $\xi_{e}$ is also
allowed to vary in addition to $N_{{\rm eff}}$ and $\eta$. In Fig.~\ref{fig:Neff_xie_eta},
we show constraints on the $\eta$--$N_{{\rm eff}}$ plane (top panel), the $N_{{\rm eff}}$--$\xi_{e}$ plane (bottom right panel) and
the $\eta$--$\xi_{e}$ plane (bottom left panel). In the figure, the fine
structure constant is fixed as $\alpha/\alpha_{0}=0.98$~(red shaded
region) and $1$~(blue contours), however the other 3 parameters,
$N_{{\rm eff}},\xi_{e}$ and $\eta$ are varied. Since we show the
constraint in a two dimensional plane, we need to project a three dimensional
constraint to the two dimensional one, which is done by fixing the redundant
parameter so that the value of $\chi^{2}$ is minimized for the parameter.
In the top panel of Fig.~\ref{fig:Neff_xie_eta}, the best-fit value of $N_{\rm eff}$ for the case with $\alpha/\alpha_0 =0.98$ is shifted slightly towards a higher ones than the standard value compared to that for $\alpha/\alpha_0 =1$. However, notice that this does not indicate that a smaller $\alpha$ prefers a slightly larger $N_{\rm eff}$. This is attributed to the fact that, in the analysis with $\xi_e$ being varied as performed in the top panel, a positively large value of $\xi_e$ is favored, which drives $N_{\rm eff}$ to be close to the standard value in the case of $\alpha/\alpha_0 =1$.  To see this in a clear manner, we show the density plot of $\xi_e$ which minimizes the value of $\chi^2$ for a given $\eta$ and $N_{\rm eff}$ with $\alpha / \alpha_0 =1$ (left) and $0.98$ (right) in Fig.~\ref{fig:Neff_eta_byxie}. 
For the case of $\alpha / \alpha_0 =1$, the allowed region lies around $N_{\rm eff} = 3.046$ and $\xi_e$ takes a positively non-zero value. On the other hand, when $\alpha / \alpha_0 =0.98$, the slightly larger value of $N_{\rm eff}$ is preferred with vanishing $\xi_e$. 
In any case, when $\xi_e$ is varied, $N_{\rm eff} = 3.046$ is well allowed in either case of $\alpha/\alpha_0 =0.98$ or $1$. 

From the constraint in the $N_{{\rm eff}}$--$\xi_{e}$ plane shown in the bottom right panel of Fig.~\ref{fig:Neff_xie_eta},
although a non-zero value of $\xi_{e}$ and a slightly higher value
of $N_{{\rm eff}}$ are preferred from the EMPRESS $Y_{p}$ for $\alpha/\alpha_{0}=1$~\cite{Matsumoto:2022tlr}, 
when $\alpha/\alpha_{0}=0.98$ is assumed, we do not need extra radiation component nor lepton asymmetry: $N_{{\rm eff}}=3.046$
and $\xi_{e}=0$ are well within 1$\sigma$ allowed region. 

This can also be seen from the bottom left panel in Fig.~\ref{fig:Neff_xie_eta}
where the constraint on the $\eta$--$\xi_{e}$ plane is shown with
$N_{{\rm eff}}$ being marginalized. One can see that $\xi_{e}$
can be zero for the case of $\alpha/\alpha_{0}=0.98$ (on the other
hand, as already known \cite{Matsumoto:2022tlr}, a non-zero positive
value of $\xi_{e}$ is preferred in the case of $\alpha/\alpha_{0}=1$).
Therefore we can conclude that the time-varying fine structure constant which is 2\% smaller than the
present value can explain the recent EMPRESS result on $Y_{p}$ without assuming non-standard $N_{\rm eff}$ and $\xi_e$.

\section{Conclusion and discussion \label{sec:conclusion}}

In this paper, we have investigated the impact of the time variation of $\alpha$ to the recent measurement of the primordial helium abundance by EMPRESS \cite{Matsumoto:2022tlr}. In the standard $\Lambda$CDM+$N_{\rm eff}$ model, the combination of the measurements of $Y_p$ by EMPRESS and $D_p$ \cite{Cooke:2017cwo} may indicate that $N_{\rm eff}$ is smaller than 3.046 and the baryon density is in a slight tension with the one obtained by Planck. However, as shown in Fig.~\ref{fig:alpha_chi2}, by assuming the value of $\alpha$ to be $2 \%$ smaller than the present value, the standard assumption of $N_{\rm eff}=3.046$ can well fit the data and the baryon density can also be consistent with the one obtained from Planck. 

Actually, if one allows the lepton asymmetry characterized by $\xi_e$ to vary, the EMPRESS result may indicate a non-zero lepton asymmetry and a slightly larger value of $N_{\rm eff}$ than the standard one \cite{Matsumoto:2022tlr}. However, as we have shown, just by assuming the time variation of $\alpha$ in the BBN epoch, one can have a good fit to the EMPRESS result, along with the measurement of D abundance~\cite{Cooke:2017cwo}, without any extra assumptions such as lepton asymmetry and non-standard value for $N_{\rm eff}$. 

Our analysis suggests that the recent measurement of $Y_p$ by EMPRESS may indicate the time variation of the fine structure constant, which would gives a significant implication on the fundamental physics, since it means that some fundamental constant could change in time. Although we have assumed that only the fine structure constant changes with time in this paper, in some theories, other constants can simultaneously be altered. It would be interesting to pursue such possibilities in the light of the EMPRESS $Y_p$ result, which is left for the future work.

\section*{Acknowledgements}We would like to thank Kazuhide~Ichikawa for sharing his BBN code with the variation of the fine structure constant. We are also grateful to Akinori~Matsumoto for the correspondence regarding the analysis in Ref \cite{Matsumoto:2022tlr}. We would also thank Masahiro~Kawasaki for a useful conversation. This work was supported by JSPS KAKENHI Grant Number 19K03874 (TT), 19K03860, 19K03865, 23K03402 (OS), MEXT KAKENHI 23H04515 (TT), and JST SPRING, Grant Number JPMJSP2119 (YT).

\clearpage 

\bibliography{BBN_me}

\end{document}